\documentclass[prb,twocolumn,secnumarabic,floatfix,amssymb,amsmath,showpacs,aps]{revtex4-1}

\usepackage{graphicx,color,epsfig,rotate}
\usepackage{times}
\usepackage{color}
\usepackage{graphicx}
\usepackage{dcolumn}
\usepackage{amssymb}
\usepackage{amsmath}
\usepackage{amsfonts}
\usepackage{bm}%
\usepackage{wrapfig}
\usepackage[colorlinks=true,linkcolor=blue,pagecolor=blue,filecolor=blue,menucolor=blue,urlcolor=blue,citecolor=blue,anchorcolor=blue]{hyperref}%
\usepackage{bm}
\usepackage{graphics}
\usepackage{graphicx}
\usepackage{color}
\usepackage{amssymb}
\usepackage{standalone}
\usepackage[toc,page]{appendix}
\usepackage{import}
\usepackage{enumerate}
\usepackage{sidecap}
\usepackage[caption=false]{subfig}

\newcommand{\lm}{\lambda}
\newcommand{\te}{\theta}
\newcommand{\Dl}{\Delta}
\newcommand{\Om}{\Omega}

\newcommand{\dl}{\delta}

\newcommand{\al}{\alpha}
\newcommand{\ep}{\varepsilon}

\newcommand{\sqh}[1]{{#1}^{\frac{3}{2}}}

\newcommand{\ovr}[2]{\langle{#1}\vert{#2}\rangle} 

\newcommand{\upw}{\uparrow}
\newcommand{\dnw}{\downarrow}

\newcommand{\lpa}[1]{\left({#1}\right)}    
\newcommand{\lca}[1]{\lbrace{#1}\rbrace}   
\newcommand{\abs}[1]{\lvert{#1}\rvert}     
\newcommand{\bra}[1]{\langle{#1}\vert}     
\newcommand{\ket}[1]{\vert{#1}\rangle}     

\newcommand{\tl}{\tilde}                   
\newcommand{\mcl}[1]{\mathcal{#1}}         
\newcommand{\tlm}[1]{\tilde{\mathcal{#1}}}  

\newcommand{\beq}{\begin{equation}}                        
\newcommand{\eeq}{\end{equation}}                          
\newcommand{\eq}[1]{\begin{equation}{#1}\end{equation}}    

\begin{document}

\title{Spin-orbit signatures in the dynamics of singlet-triplet qubits in double quantum dots}

\author{Juan E. Rolon}
\email{rolon@email.unc.edu}
\affiliation{Department of Physics and Astronomy, University of North Carolina, Chapel Hill, North Carolina, 27599-3255, USA}
\affiliation{Centro de Investigaci\'{o}n Cient\'{i}fica y de Educaci\'{o}n Superior de Ensenada, Apartado Postal 360, Ensenada, Baja California, 22800, M\'{e}xico}

\author{Ernesto Cota}
\email{ernesto@cnyn.unam.mx}
\affiliation{Centro de Nanociencias y Nanotecnolo´g\'ia, Universidad Nacional Aut\'onoma de M\'exico, Apartado Postal 14, Ensenada, Baja California, 22800 M\'exico}

\author{Sergio E. Ulloa}
\email{ulloa@ohio.edu}
\affiliation{Department of Physics and Astronomy, and Nanoscale and Quantum Phenomena Institute, Ohio University, Athens, Ohio, 45701-2979, USA}

\date{\today}

\begin{abstract}
We characterize numerically and analytically the signatures of the spin-orbit interaction in a two-electron GaAs double quantum dot in the presence of an external magnetic field. In particular, we obtain the return probability of the singlet state by simulating Landau-Zener voltage detuning sweeps which traverse the singlet-triplet ($S-T_+$) resonance. Our results indicate that non-spin-conserving interdot tunneling processes arising from the spin-orbit interaction have well defined signatures. These allow direct access to the spin-orbit interaction scales and are characterized by a frequency shift and Fourier amplitude modulation of the Rabi flopping dynamics of the singlet-triplet qubits $S-T_0$ and $S-T_+$. By applying the Bloch-Feshbach projection formalism, we demonstrate analytically that the aforementioned effects originate from the interplay between spin-orbit interaction and processes driven by the hyperfine interaction between the electron spins and those of the GaAs nuclei.
\end{abstract}

\maketitle

\section{Introduction}
\label{SectionIntro}
Electron spins in double quantum dots (DQDs) in the spin-blockade regime, with one or two electrons confined in each dot, have been proposed as qubits for quantum computing implementations at the nanoscale.~\cite{Loss-DiVincenzo} From both theoretical and experimental points of view, the coherent control of these spin qubits has become a subject of great interest in the condensed matter and quantum information communities. In particular, there is considerable emphasis on the study of their coherent dynamics and the limiting mechanisms of spin coherence in the presence of hyperfine interactions with the ensemble of nuclear spins of the DQD host material.~\cite{Elzerman,Johnson,Trauzettel,Nowack,Bluhm}

Singlet-triplet qubits have been implemented in DQDs built in either GaAs or Si/Ge hosts. They are formed by the singlet state $S(1,1)$---with one electron in each dot---and either of the triplet states $T_0$ $(m=0)$ or $T_+$ $(m=1)$. Universal control has been demonstrated using quantum state tomography obtained by control pulses (or sweeps) of the voltage difference (detuning) between the QDs.~\cite{Petta, Eriksson, Maune, Veldhorst} Indeed, such detuning sweeps allow the implementation of qubit rotations around a single axis of the Bloch sphere ($S-T_0$ qubit exchange gates) resulting from the dynamic variation of the two-electron exchange interaction. Furthermore, in combination with dynamic nuclear polarization (DNP) techniques, which rely on traversing the $S-T_+$ qubit resonance via Landau-Zener (LZ) detuning sweeps, a sustained magnetic hyperfine field gradient between the QDs can be created for times longer than 30 min, thus allowing coherent rotations around two axes of the Bloch sphere, an essential requisite for the implementation of universal quantum gates.~\cite{YacobyFoletti,Coish-Loss}

An important question concerning the limiting factors to DNP efficiency when traversing the $S-T_+$ resonance, is whether the spin-orbit interaction (SOI) could influence DNP transfer from the electron to the nuclear spins, and consequently affect the fidelity of the singlet-triplet qubits.~\cite{Stepanenko,Burkard} In particular, a quenching of DNP has been observed when SOI exceeds the hyperfine interaction, preventing an increase in the spin decoherence time in GaAs quantum dots.~\cite{Nichol} Recently, interferometry experiments have been able to probe the fast dynamics associated with $S-T_+$ transitions by correlating the outcomes of an ensemble of individual single shot measurements of the qubit state after LZ transitions.~\cite{Dickel}

In this work, we provide further insights into SOI signatures that could be probed by one-shot readout experiments measuring the singlet state return probability following a rapid LZ detuning sweep traversing the $S-T_+$ resonance. We perform both numerical and analytical calculations based on a realistic model, accounting for the dynamics of the lifting of the spin-blockade regime via SOI. We show that the signatures of SOI are manifested in a frequency shift near the vicinity of the $S-T_+$ resonance, and detuning-dependent modulation of the Fourier amplitudes corresponding to transitions between the singlet state $S(1,1)$ and the triplets $T_0$ and $T_+$. The present analysis could become useful to experimentalists in search of direct measurements of SOI in a given system without the need for correlation measurements.~\cite{Dickel}

This paper is organized as follows. Section \ref{SectionModel} introduces a realistic Hamiltonian model of the DQD subjected to a voltage detuning and an external magnetic field, and takes into account interdot spin conserving and non-spin-conserving tunneling processes originating from the interplay between SOI and the nuclear hyperfine interaction. Section \ref{LACSPopulationSpectrum} discusses the level-anticrossings in the spectrum of the system and the state mixing effects resulting from voltage detuning, Zeeman splitting, SOI and the hyperfine magnetic field. In Sec.\ \ref{Signatures}, we discuss how SOI signatures can be detected via singlet return probability maps derived from LZ sweeps across the $S-T_+$ resonance. In Sec.\ \ref{FourierAmplitudes} we present numerical and analytical results for potential SOI signatures in the Fourier amplitudes and frequency shifts associated with $S-T_0$ and $S-T_+$ transitions. Section \ref{Noise} discusses the effects of electrical noise and nuclear hyperfine field fluctuations. Finally, we present a summary and discussion of our results in Sec.\ \ref{Conclusion}.

\section{Model}
\label{SectionModel}

The system under consideration consists of a gate-defined GaAs DQD having a total occupation of two electrons. The charge state of the DQD and the spatial separation of the electrons is determined by a detuning parameter $\ep$, which controls the relative electrostatic potential of the quantum dot pair. In the limit of a small perpendicular applied magnetic field, the relevant occupied states are singlets, $\ket{S(0,2)} = (\ket{\upw_R  \dnw_R} - \ket{\dnw_R \upw_R})/\sqrt{2}$ for $\ep \gg 0$ and $\ket{S(1,1)}=(\ket{\upw_L  \dnw_R} - \ket{\dnw_L \upw_R})/\sqrt{2}$ for $\ep \ll 0$, where $(n_L,n_R)$ denotes the occupation of the left (L) and right (R) QDs, respectively. The Hamiltonian that describes the coupling between the two singlets is given by
\begin{equation}
\mcl{H}_{t_0} = -\ep\ket{S(0,2)}\bra{S(0,2)} + t_0 \ket{S(1,1)}\bra{S(0,2)} + \text{H.c.}
\label{Htunnel}
\end{equation}
where $t_0$ is the interdot spin-conserving tunneling strength. In the far negative detuning regime, $\ep \ll 0$, the ground state of the system becomes the $\ket{S(2,0)}$ singlet; this is a far off-resonant state which has a negligible effect on the system dynamics and is henceforth ignored in the model. The separation of the electronic wave functions causes $S(1,1)$ (henceforth denoted by $S$) to be nearly degenerate with the triplet states $T_m$ $(m={1,0,-1})$, i.e. $\ket{T_+} = \ket{\upw_L \upw_R}$, $\ket{T_0}=(\ket{\upw_L  \dnw_R} + \ket{\dnw_L \upw_R})/\sqrt{2}$ and $\ket{T_-} = \ket{\dnw_L \dnw_R}$. The states $S$ and $T_0$ have zero spin angular momentum component along the z-axis, and can be used as a suitable computational basis for spin qubits in DQDs~\cite{YacobyFoletti}, whereas the degeneracy with the triplet states $T_+$ and $T_-$ is lifted by the application of an external magnetic field $\vec{B}_{\text{ext}}$, which allows the qubit to be controllable by the triplet states and the outgoing singlet $S(0,2)$. Furthermore, the interaction of the electron spins with the nuclear magnetic field of the host material can be harnessed to control the $S-T_0$ qubit dynamics by the internally created magnetic field gradient across the DQD structure, $d\vec{B} = (\vec{B}_{\text{nuc,L}} - \vec{B}_{\text{nuc,R}})/2$. The hyperfine interaction with $d\vec{B}$ couples the singlet $S$ to the triplet states via non-spin-conserving transitions. The Hamiltonian which describes the interaction between electron spins, and their interaction with the hyperfine and external magnetic fields is given by~\cite{Taylor}
\begin{equation}
\mcl{H}_{\rm hf} = \vec{B} \cdot (\vec{S}_L + \vec{S}_R) + d\vec{B} \cdot (\vec{S}_L - \vec{S}_R) \, ,
\label{HHyperfine}
\end{equation}
where $\vec{B} = \vec{B}_{\text{ext}} + \vec{B}_{\text{nuc}}$ is the total magnetic field, $\vec{B}_{\text{nuc}} = (\vec{B}_{\text{nuc,L}} + \vec{B}_{\text{nuc,R}})/2$, the mean nuclear magnetic field, and $\vec{S}_L$, $\vec{S}_R$ the spins in the left and right dots, respectively. We assume $g^*\mu_B = 1$, where $g^*$ is the electron g-factor in GaAs, and $\mu_B$ is the Bohr magneton; we write all magnetic and hyperfine couplings using units of energy.

\begin{figure}[t]
\includegraphics[width=0.8 \columnwidth]{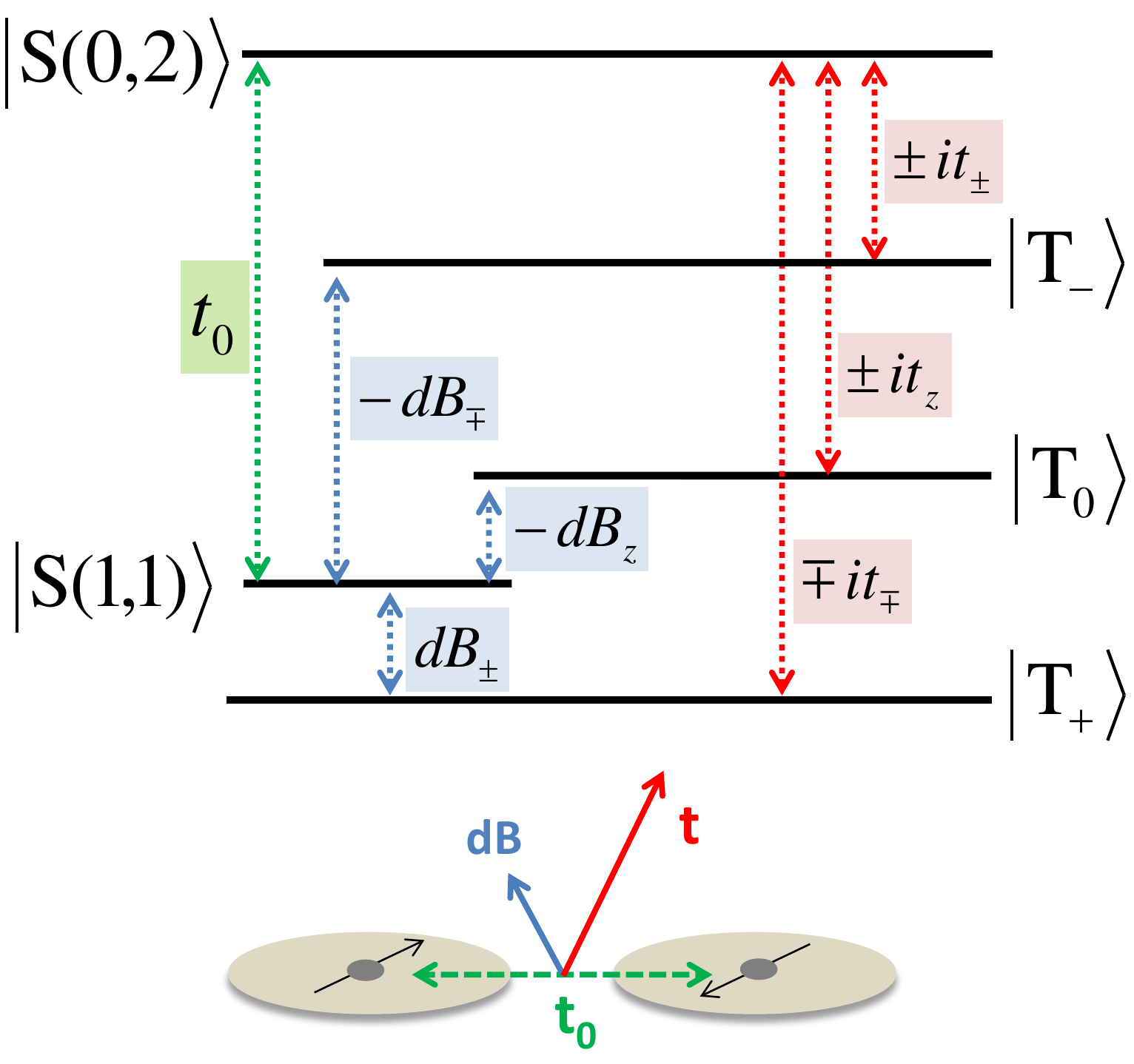}
\caption{(Color online) Energy level diagram corresponding to the Hamiltonian in Eq.\ \ref{BareHamiltonian}. Green arrows indicate spin-preserving charge tunneling-mediated transitions, while blue and red arrows, indicate hyperfine and spin-orbit spin-flip transitions, respectively. The bottom diagram illustrates the DQD system and the vector nature of the competing nuclear magnetic field gradient $d\vec{B}$ and SOI non-spin conserving interdot tunneling coupling $\vec{t}$.}
\label{BareLevelDiagram}
\end{figure}

In this system, SOI induces non-spin-conserving tunneling processes for electrons that couple the singlet $S(0,2)$ to the triplet states. This interaction, arising from intrinsic electric fields induced by structural asymmetries,~\cite{DanonNazarov,Romano, Mircea, Rashba} can lift the spin-blockade regime and provide a competing mechanism to the hyperfine-mediated electron spin flips involved in DNP. The Hamiltonian $\mcl{H}_{t_{SO}}$ associated with SOI tunneling processes can be written in the basis of orthonormal unpolarized states $\ket{\vec{T}} = \lca{\ket{T_x}, \ket{T_y}, \ket{T_z}}$ given by,~\cite{DanonNazarov}
\begin{eqnarray}
\ket{T_x} &=& \frac{1}{\sqrt{2}}(\ket{T_+} - \ket{T_-}) \\ \nonumber
\ket{T_y} &=& \frac{i}{\sqrt{2}}(\ket{T_+} + \ket{T_-}) \\ \nonumber
\ket{T_z} &=& \ket{T_0} \, ,
\end{eqnarray}
such that
\begin{equation}
\mcl{H}_{t_{SO}} = i\vec{t}\cdot\ket{\vec{T}}\bra{S(0,2)} + \text{H.c.} \, ,
\end{equation}
where $\vec{t} = \lca{t_x, t_y, t_z}$ is a real vector that defines the degree of spin state mixing due to SOI.
Accordingly, the total Hamiltonian, $\mcl{H}= \mcl{H}_{t_0} + \mcl{H}_{t_{SO}}+\mcl{H}_{\rm hf}$ describes charge tunneling transitions in the DQD and the interplay between hyperfine and spin-orbit interactions. Thus, in the singlet-triplet basis $\lca{\ket{S}, \ket{T_+}, \ket{T_0}, \ket{T_-}, \ket{S(0,2)} }$ one obtains (with $\ket{S}=\ket{S(1,1)}$),
\begin{equation}
\mcl{H } = \left(
  \begin{array}{ccccc}
    0 & dB_+ & -dB_z & -dB_- & t_0 \\
    dB_- & B_z & 0 & 0 & -it_- \\
    -dB_z & 0 & 0 & 0 & it_z \\
    -dB_+ & 0 & 0 & -B_z & it_+ \\
    t_0 & it_+ & -it_z & -it_- & -\ep \\
  \end{array}
\right) \, ,
\label{BareHamiltonian}
\end{equation}
where $dB_{\pm} = (dB_x \pm idB_y)/\sqrt{2}$ and $t_{\pm} = (t_x \pm it_y)/\sqrt{2}$ represent the couplings that induce spin-flip processes via the hyperfine field gradient and SOI non-spin-conserving tunneling, respectively; Fig.\ \ref{BareLevelDiagram} illustrates the different processes.

The contribution of both singlet states to the electron exchange energy, and their interplay with the spin-flip dynamics, is better evidenced after introducing the following change of basis,~\cite{Taylor}
\begin{equation}
\left(
  \begin{array}{c}
    \ket{\tl{S}} \\
    \ket{\tl{G}} \\
  \end{array}
\right) = \left(
  \begin{array}{cc}
    \cos{\theta} & \sin{\theta} \\
    -\sin{\theta} & \cos{\theta} \\
  \end{array}
\right) \left(
  \begin{array}{c}
    \ket{S} \\
    \ket{S(0,2)} \\
  \end{array}
\right) \, ,
\label{AdiabaticTransformation}
\end{equation}
where $\theta = \arctan{(\frac{-J(\ep)}{t_0})}$ and the coherent exchange energy of the electrons is given by $J(\ep) = \frac{1}{2}(\ep + \sqrt{\ep^2 + 4t_0^2})$, and amounts to the energy gap between the singlet $S$ and the triplet state $T_0$ in the absence of any other perturbations but charge tunneling. As such, for $\ep \ll 0$, $\tl{S} \rightarrow S$, $\tl{G} \rightarrow S(0,2)$ and for $\ep \gg 0$, $\tl{S} \rightarrow S(0,2)$, $\tl{G} \rightarrow S$, where states $\tl{S}$ and $\tl{G}$ are often referred as the {\it hybridized singlet states}~\cite{Burkard, Nichol}.  The Hamiltonian in the adiabatic basis $\lca{\ket{\tl{S}}, \ket{T_+}, \ket{T_0}, \ket{T_-} \ket{\tl{G}}}$ transforms to
\begin{equation}
\mcl{\tl{H}} = \left(
  \begin{array}{ccccc}
    -J(\ep)   & \Pi_{+}   & \Pi_z       & \Pi_{-}   & 0          \\
    \Pi_{+}^* & B_z       & 0           & 0         & \Om_{-}    \\
    \Pi_z^{*} & 0         & 0           & 0         & \Om_{z}    \\
    \Pi_{-}^* & 0         & 0           & -B_z      & \Om_{+}    \\
     0        & \Om_{-}^* & \Om_{z}^{*} & \Om_{+}^* & J(\ep)-\ep \\
  \end{array}
\right) \, ,
\label{TildeHamiltonian}
\end{equation}
where
\begin{eqnarray}
\Pi_{\pm} & = & \mp dB_{\pm}\cos{\te} \pm i t_{\pm}\sin{\te}    \label{PIPlusMinus} \\
\Om_{\pm} & = & \pm dB_{\pm} \sin{\te} \pm i t_{\pm} \cos{\te}  \label{OmegaPlusMinus} \\
\Pi_{z}   & = & dB_z\cos{\te} + i t_z\sin{\te}                  \label{PIZeta} \\
\Om_{z}   & = & dB_z\sin{\te} - i t_z \cos{\te}                 \label{OmegaZeta} \, .
\end{eqnarray}
The Hamiltonian off-diagonal matrix elements in Eqs.\ \ref{PIPlusMinus}-\ref{OmegaZeta} characterize the coupling of both hybridized singlets to the triplet states, and the competition of spin-flips induced by the hyperfine field gradient and SOI assisted tunneling transitions.

The population dynamics of the singlet and triplet states is obtained by solving the master equation, $\dot{\rho} = (-i/\hbar)[\tl{\mcl{H}},\rho]$, within the scope of the quasi-static approximation: we assume that the dephasing by hyperfine interactions with the nuclear spin bath and associated spin relaxation takes place in a time-scale ($\sim \mu s$) much longer than the time-span ($\sim$ tens of ns) of the detuning sweeps, $\ep(t)$, which are typically implemented in experiments. This allows us to consider essentially a static nuclear magnetic field over the time-span of the detuning sweeps. We discuss later, in Sec.\ \ref{Noise}, the effects of hyperfine field fluctuations and electrical noise over the course of collecting data over typical detuning sweep repetitions. For simplicity, and without loss of insight, the results presented here assume isotropic components of both the nuclear magnetic field gradient and the spin-orbit tunneling vector, i.e. $\abs{d\vec{B}} = \sqrt{3}dB$ and $\abs{\vec{t}} = \sqrt{3}t_{SO}$.

\section{Level anticrossing spectrum and singlet-triplet dynamics}
\label{LACSPopulationSpectrum}

\begin{figure}[h]
\includegraphics[width=1.0 \columnwidth]{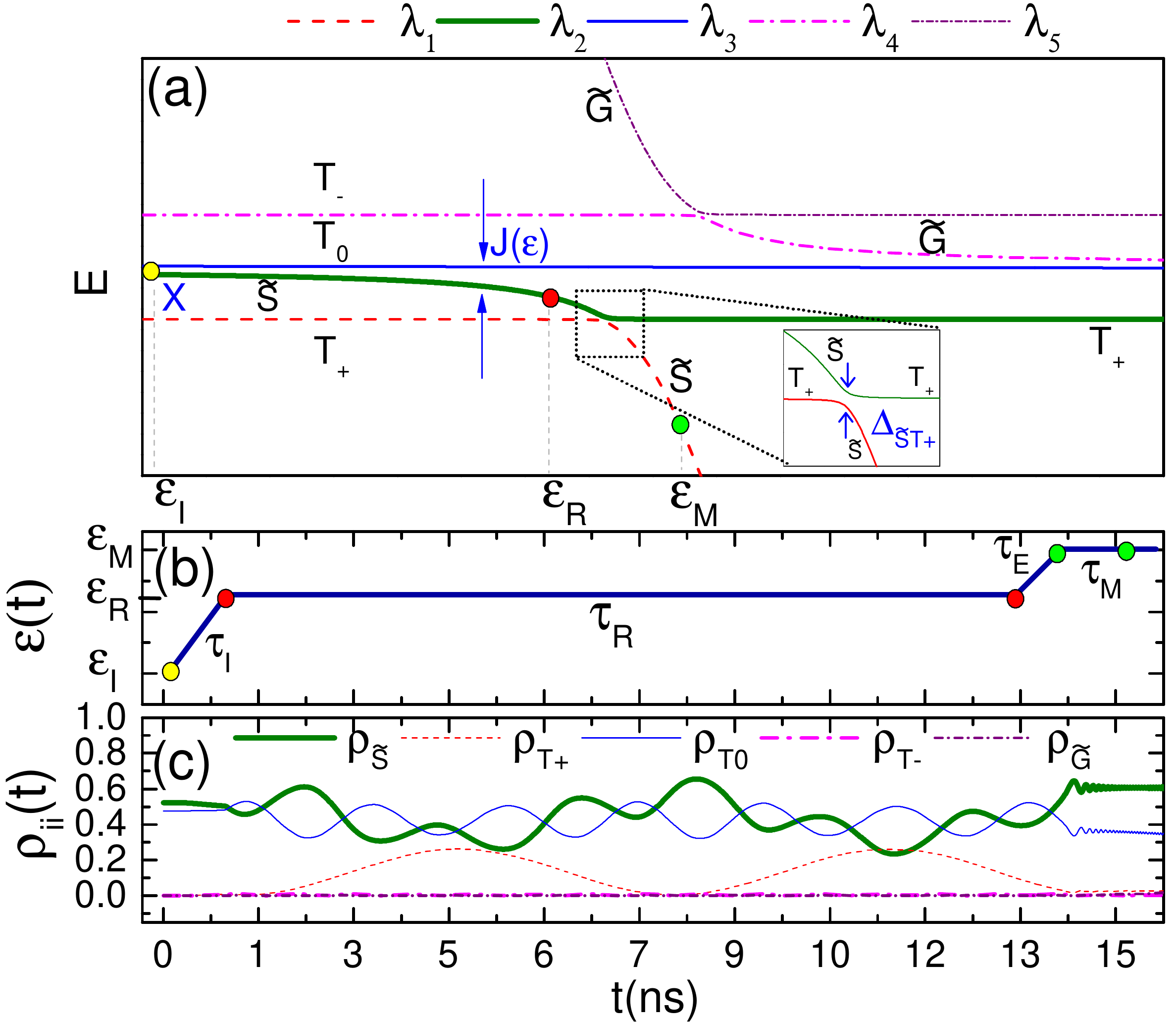}
\caption{(Color online) (a) Characteristic eigenvalue spectrum $\lca{\lm_i(\ep)}$ corresponding to the Hamiltonian in Eq.\ \ref{TildeHamiltonian}. The yellow dot indicates the initialization stage in the eigenstate $\ket{X} \equiv \ket{\lm_2} \simeq (\ket{\tl{S}} + \ket{T_0})/\sqrt{2}$ for a detuning $\ep_I$. The red dot indicates the detuning value, $\ep_R < 0$, at which the system is allowed to evolve during a residence time $\tau_R$. The green dot indicates the detuning, $\ep_M > 0$, at which the singlet return probability $P_{\tl{S}}$ is measured. $J(\ep)$ is the exchange energy splitting. After initialization at $\ep_I \ll 0$ and residence at $\ep_R$,  $\tau_I$ and $\tau_E$ represent sudden detuning pulses of $\sim 1\text{ns}$ duration, respectively. The inset shows the singlet-triplet anticrossing splitting, $\Delta_{ST_+}$, mediated by the hyperfine and spin-orbit coupling. (b) Sequence of detuning sweeps used to control the DQD state dynamics. The vertex $(\tau_I, \ep_I)$ corresponds to initialization, while $(\tau_R, \ep_R)$ and $(\tau_M, \ep_M)$ correspond to residence and measurement control stages, respectively. (c) State dynamics according to the detuning control scheme in (b).}
\label{GenericSpectrumPulse}
\end{figure}

Figure \ref{GenericSpectrumPulse}(a) shows the level spectrum corresponding to the Hamiltonian in Eq.\ \ref{TildeHamiltonian}, indicating the behavior of the different state mixing resonances, and their dependence on the interdot coupling parameters and applied energy detuning $\ep$.
The eigenvalues, $\lm_i$, exhibit a series of avoided crossings dominated here by spin conserving tunneling, $t_0$, which produces the hybridized singlets $\tl{S}$ and $\tl{G}$; this anticrossing is set to occur at $\ep=0$. Two singlet-triplet anticrossings appear, $\tl{S}-T_+$ at $\ep < 0$, and $\tl{G}-T_-$ at $\ep >0$, respectively. Notice that $t_0 > \Dl_{\tl{S}T_0} \sim dB, t_{SO}$ in this diagram, which is typical in real DQD systems. For $\ep \ll 0$, the triplet $T_+$ becomes the ground state while $T_0$ and $\tl{S}$ approach a degeneracy point. In this regime, the first excited eigenstate of the system is approximately $\ket{X} \sim (\ket{\tl{S}}+\ket{T_0})/\sqrt{2}$. Similarly, for $\ep \gg 0$ the excited eigenstate of the system approximates $\ket{X^\prime} \sim (\ket{\tl{G}}+\ket{T_0})/\sqrt{2}$. In the presence of spin-mixing terms, the energy gap between the triplet state $T_0$ and the hybridized singlet states $\tl{S}$ and $\tl{G}$ is given by $-J(\ep)$ and $J(\ep)-\ep$, respectively. The regimes of interest in this work correspond to the qubit subspaces defined in the vicinity of the $\tl{S}-T_0$ degeneracy point and the $\tl{S}-T_+$ resonance, where SOI signatures are more important and can be probed through an analysis of the Fourier amplitudes and frequency shifts of the (Rabi oscillations) populations of the different DQD states, as we will see.

The dynamics of the singlet state $\tl{S}$ and triplet states, $T_0$ and $T_+$, is controlled through the sequence of LZ detuning sweeps shown in Fig.\ \ref{GenericSpectrumPulse}(b), which are similar to those implemented in experiments.~\cite{Petta,YacobyFoletti} The system is initialized in the eigenstate $\ket{X}$ at $\ep \ll 0$, where, $\tl{S}-T_0$ transitions are mainly driven by the axial component of the nuclear magnetic field gradient, $dB_z$. The detuning is then subjected to a rapid sweep of duration $\tau_I \sim 1$ns, which drives the system near both the $\tl{S}-T_+$ and $\tl{S}-\tl{G}$ avoided crossings (this point in red is labelled by $\ep_R$ in Fig.\ \ref{GenericSpectrumPulse}(a)). Notice that the exchange energy, $J(\ep)$, changes rapidly during this stage. The system is allowed now to evolve during a residence time $\tau_R\sim1-60$ns. Figure \ref{GenericSpectrumPulse}(c) shows the DQD state dynamics when the system evolves during one of these detuning sweep sequences. The fast Rabi oscillations between states $\tl{S}$ and $T_0$ occur with a frequency, $(\lm_3-\lm_2)/\hbar \simeq f_{\tl{S}-T_0} \simeq J(\ep)/\hbar$. The amplitude of these oscillations follows an envelope that oscillates in phase with the amplitude of the triplet $T_+$; as the system is near the $\tl{S}-T_+$ resonance, the corresponding frequency is given by $(\lm_2-\lm_1)/\hbar \simeq \Dl_{\tl{S}-T_+}/\hbar < f_{\tl{S}-T_0}$. As shown in Sec.\ref{FourierAmplitudes} and Appendix \ref{FeshbachProjection}, the Rabi frequencies and oscillation amplitudes are strongly modified by the interplay of exchange and non-spin conserving processes due to SOI and the hyperfine interaction. In this regime, the hybridized singlets, $\tl{S}$ and $\tl{G}$, couple to all triplet states $T_m$, and a competition between the components of the hyperfine field gradient $(dB_x, dB_y)$ and SOI $(t_x, t_y)$ has a strong effect on the system dynamics.

In order to probe the corresponding signatures, an additional rapid detuning pulse of duration $\tau_E\sim 1$ns traverses the $\tl{S}-T_+$ resonance and drives the system beyond the charge transition anticrossing until reaching the detuning value $\ep_M$. Here, the hybridized singlet, $\tl{S}$, has a significant component along the outgoing singlet, $S(0,2)$, whose charge state is typically measured using a quantum point contact. The probability, $P_{\tl{S}}$, of recovering the singlet state $S$ is obtained by computing the average population $\rho_{\tl{S}}$ over the measuring time interval $t_f - t_i = \tau_M\sim 1$ns, while the system evolves at fixed detuning $\ep_M$,
\begin{equation}
P_{\tl{S}} = \frac{1}{\tau_M} \int_{t_i}^{t_f} \rho_{\tl{S}}(t) dt \, .
\label{ReturnProbability}
\end{equation}
As shown below, the signatures resulting from the interplay of SOI and the hyperfine interaction emerge clearly in the behavior of $P_{\tl{S}}$, as the residence time $\tau_R$ is varied.

\begin{figure*}[t]

\subfloat{%
  \includegraphics[clip,width=0.4\textwidth]{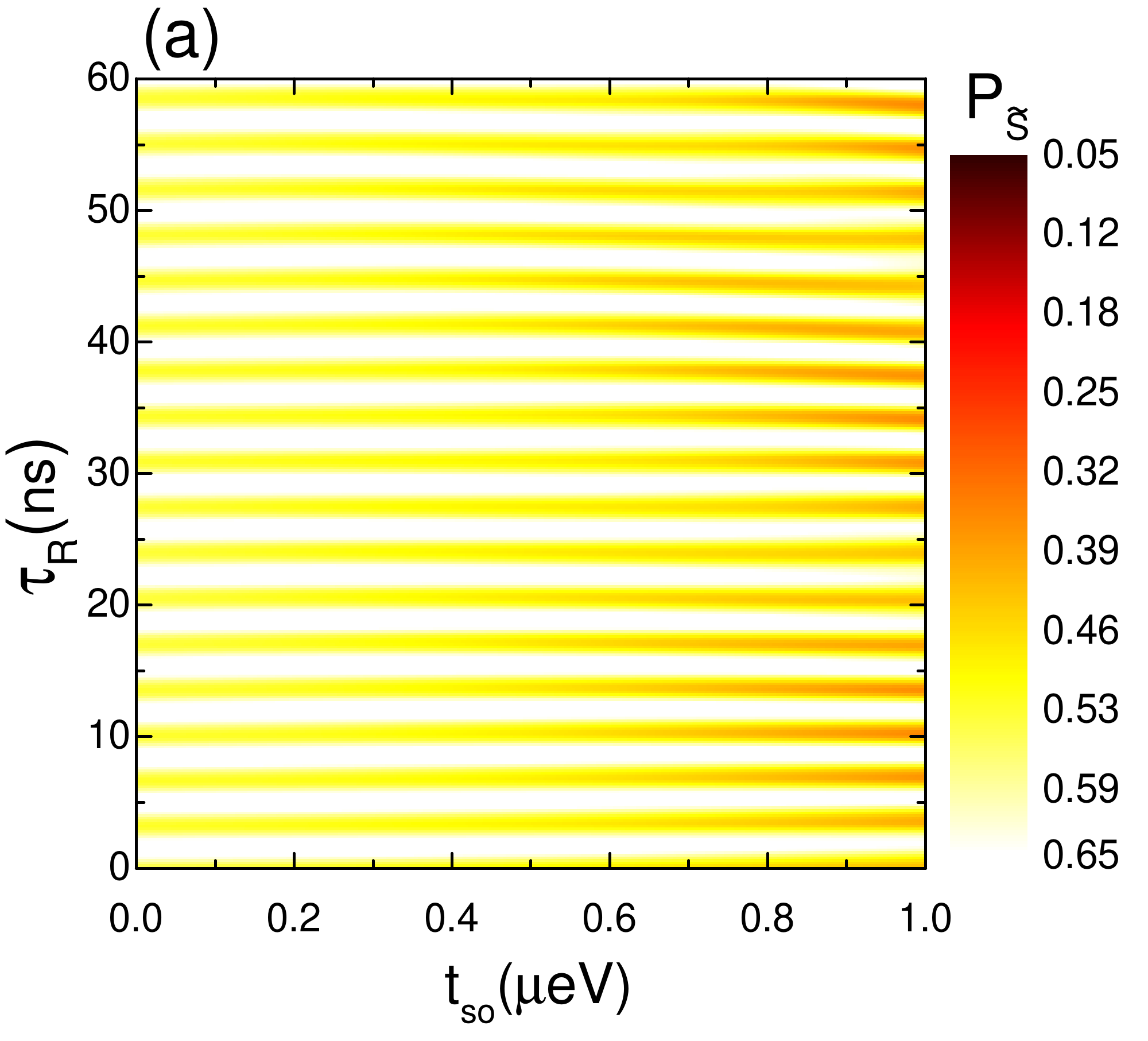}%
}
\subfloat{%
  \includegraphics[clip,width=0.4\textwidth]{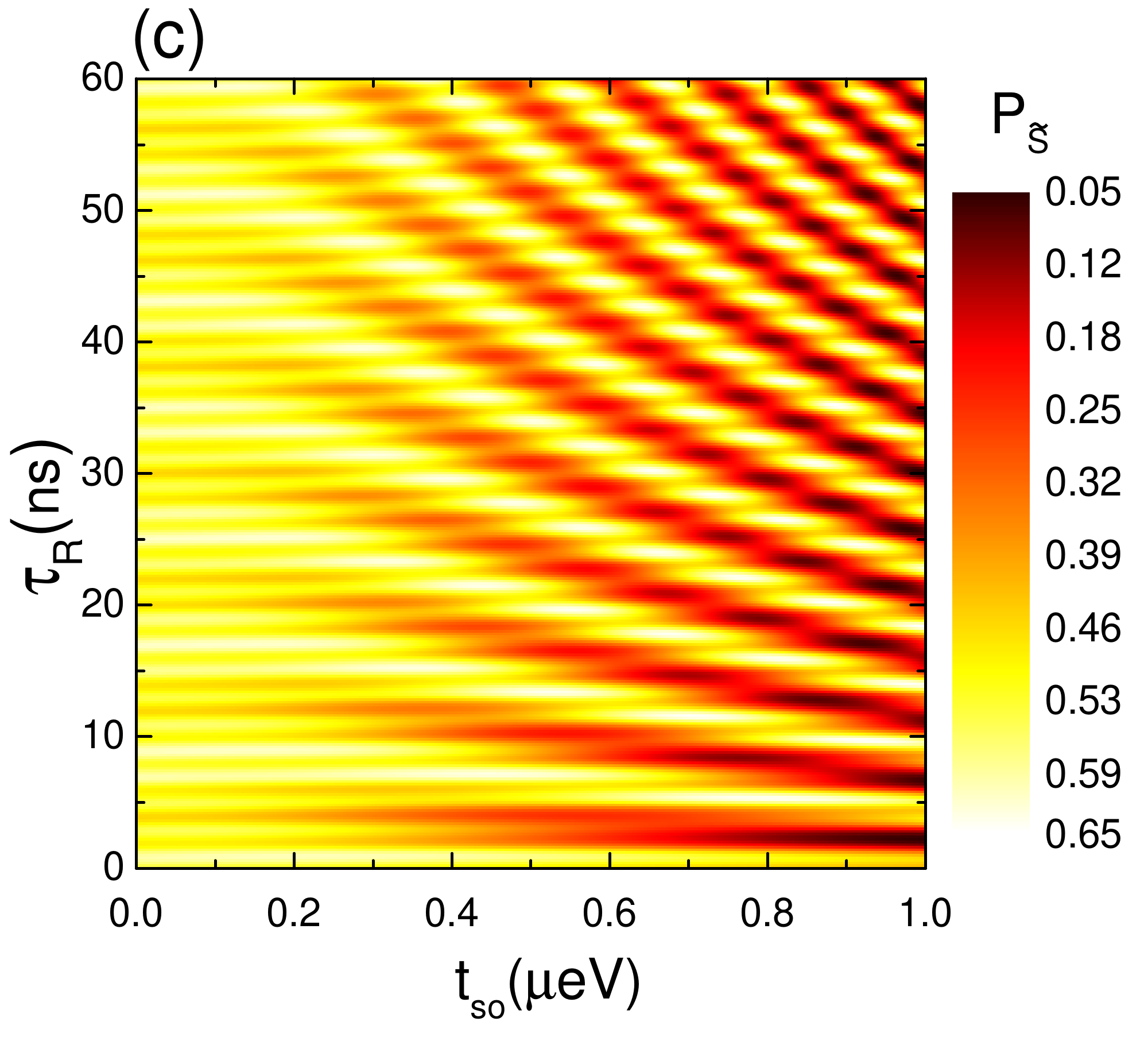}%
}

\subfloat{%
  \includegraphics[clip,width=0.4\textwidth]{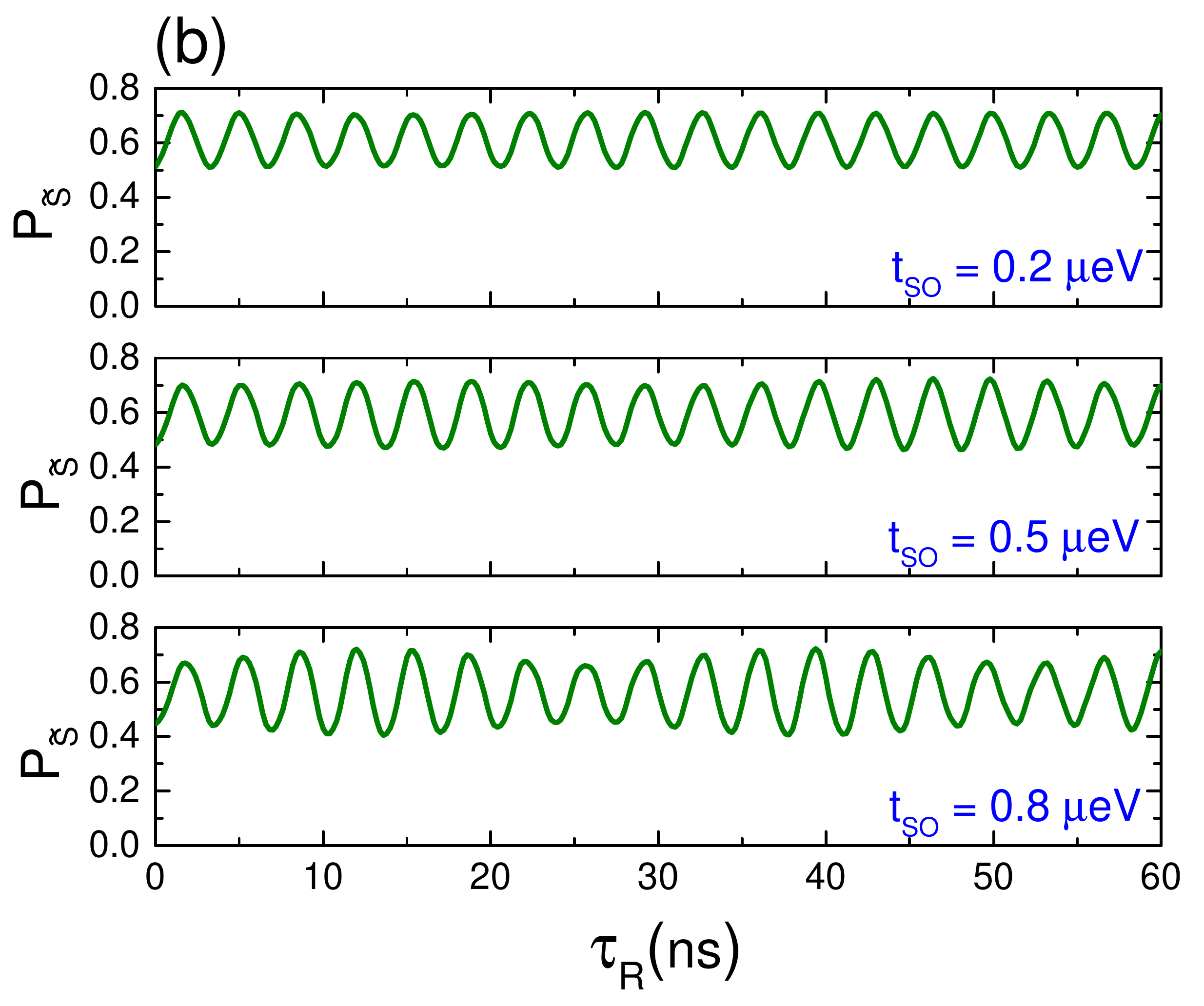}%
}
\subfloat{%
  \includegraphics[clip,width=0.4\textwidth]{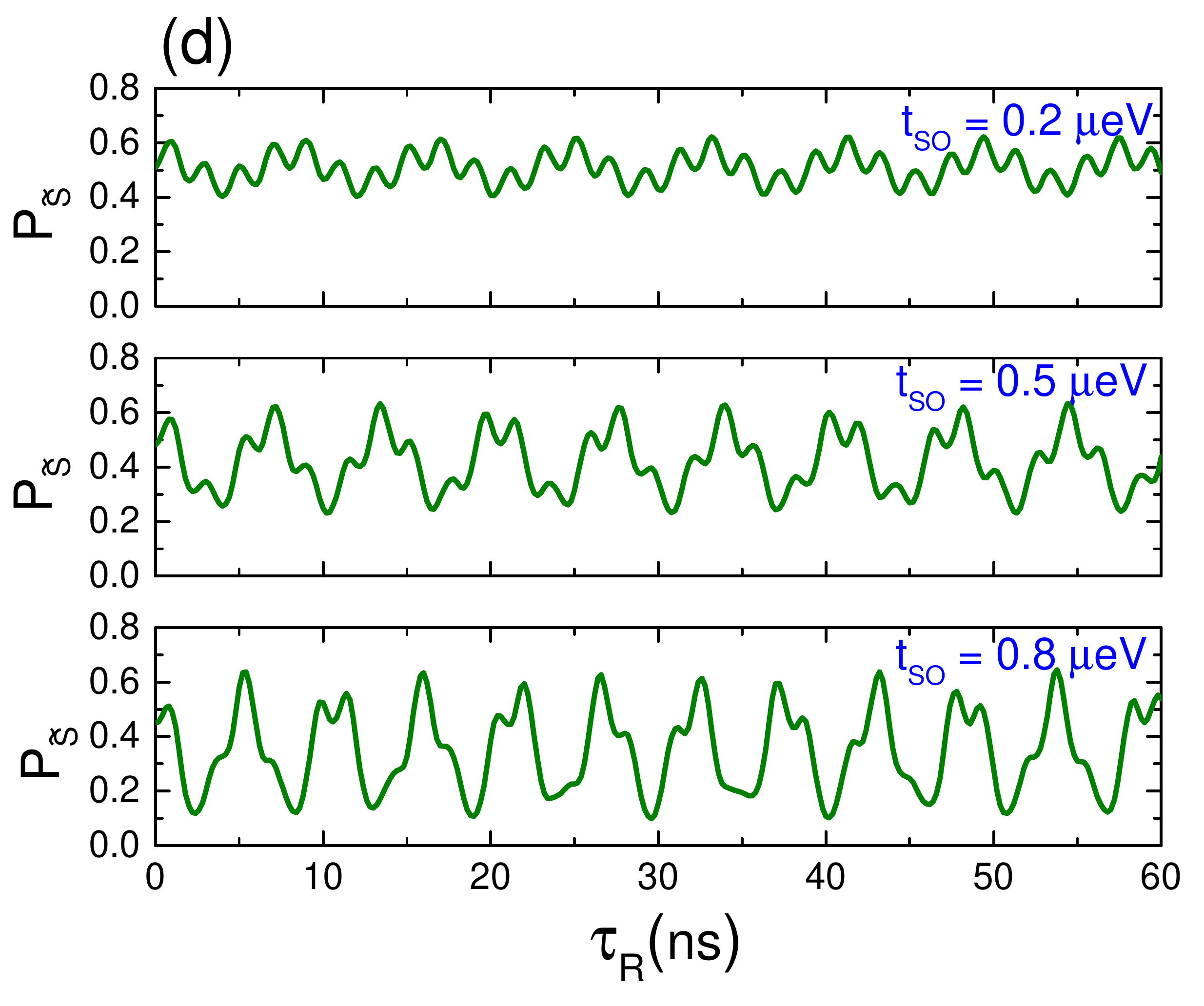}%
}

\caption{(Color online) SOI signatures associated with the singlet return probability $P_{\tl{S}}$. (a) and (c) Singlet return probability map as function of spin-orbit coupling, $t_{SO}$, and residence time, $\tau_R$, for residence detuning values of $\ep_R = -20\mu$eV and $\ep_R = -10\mu$eV, respectively. (b) and (d) Singlet return probability as function of $\tau_R$ for $t_{SO}= 0.2, 0.5, 0.8 \mu$eV, corresponding to cuts along the vertical axis in panels (a) and (c), respectively.}
\label{SpinOrbitSignatures}
\end{figure*}

\section{SOI signatures on the singlet return probability}
\label{Signatures}

The signatures of SOI in $P_{\tl{S}}$ resulting from the detuning sweeps described above are illustrated in Fig.\ \ref{SpinOrbitSignatures}, which shows a pair of maps of the singlet return probability $P_{\tl{S}}$ as function of the residence time $\tau_R$, non-spin conserving tunneling strength $t_{SO}$, and different $\ep_R$ values. Each point in the map corresponds to a single shot realization of the pulse sequence described in Fig.\ \ref{GenericSpectrumPulse}(b). The following parameters (typical of experimental DQD systems) were used in our simulations: spin-conserving tunneling coupling $t_0=5\mu\text{eV}$, Zeeman splitting $E_Z = g\mu_B B = 2.5\mu\text{eV}$, nuclear magnetic field gradient $dB=0.125\mu\text{eV}$, see Refs.\ [\onlinecite{YacobyFoletti,DanonNazarov}]. In both maps, initialization occurs at $\ep_I =-2000\mu\text{eV}$, and measurement at $\ep_M=+90\mu\text{eV}$. Figure \ref{SpinOrbitSignatures}(a) shows the behavior for $\ep_R = -20\mu\text{eV}$, where $P_{\tl{S}}$ exhibits oscillations with a characteristic period of $3.4\text{ns} \simeq \hbar/1.2\mu\text{eV}$, over the entire range of $t_{SO}$. As $t_{SO}$ increases, however, the periodic oscillations occur accompanied with an envelope modulation due to the presence of an additional frequency. The modulation is perhaps more evident in Fig.\ \ref{SpinOrbitSignatures}(b), and the corresponding cross section of the map in Fig.\ \ref{SpinOrbitSignatures}(a), especially for $t_{SO}\simeq 0.8\mu\text{eV}$.

Figure \ref{SpinOrbitSignatures}(c) shows the $P_{\tl{S}}$ map for a residence detuning $\ep_R = -10\mu\text{eV}$, i.e. much closer to the $\tl{S}-T_+$ resonance. Here, as spin-flip processes are more pronounced, the behavior changes dramatically. We notice that the oscillations exhibit more pronounced minima (darker colors) towards the right-hand side of the map, while a strong frequency shift and differentiated pattern is evident in the increasing number of alternating dark and bright contour regions towards higher $t_{SO}$ values. Interestingly, somewhat sudden phase shifts of the $P_{\tl{S}}$ oscillations are observed for increasing $t_{SO}$. The map cross section in Fig.\ \ref{SpinOrbitSignatures}(d) corroborates the aforementioned behavior, clearly exhibiting additional frequency components in the signal envelope, which appear better resolved as $t_{SO}$ is increased.

\section{Bloch-Feshbach projection and SOI signatures on the Fourier amplitudes of $P_{\tl{S}}$ oscillations}
\label{FourierAmplitudes}

The behavior of the amplitudes and frequency shifts associated with the oscillatory behavior of $P_{\tl{S}}$ as function of $\tau_{R}$ can be explained analytically using the Bloch-Feshbach projection method (see Appendix \ref{FeshbachProjection}). By projecting out the hybridized singlet $\tl{G}$ and triplet state $T_-$, we obtain an effective Hamiltonian, $\mcl{H}_{\tl{S}T}$, which describes the system dynamics within the subspace spanned by the basis $\lca{\ket{\tl{S}}, \ket{T_+}, \ket{T_0}}$. These states are the ones relevant in the dynamical processes involving the interplay of the hyperfine field and SOI near the $\tl{S}-T_+$ resonance.

\begin{figure*}[ht]
    \centering
    \includegraphics[width=0.8 \textwidth]{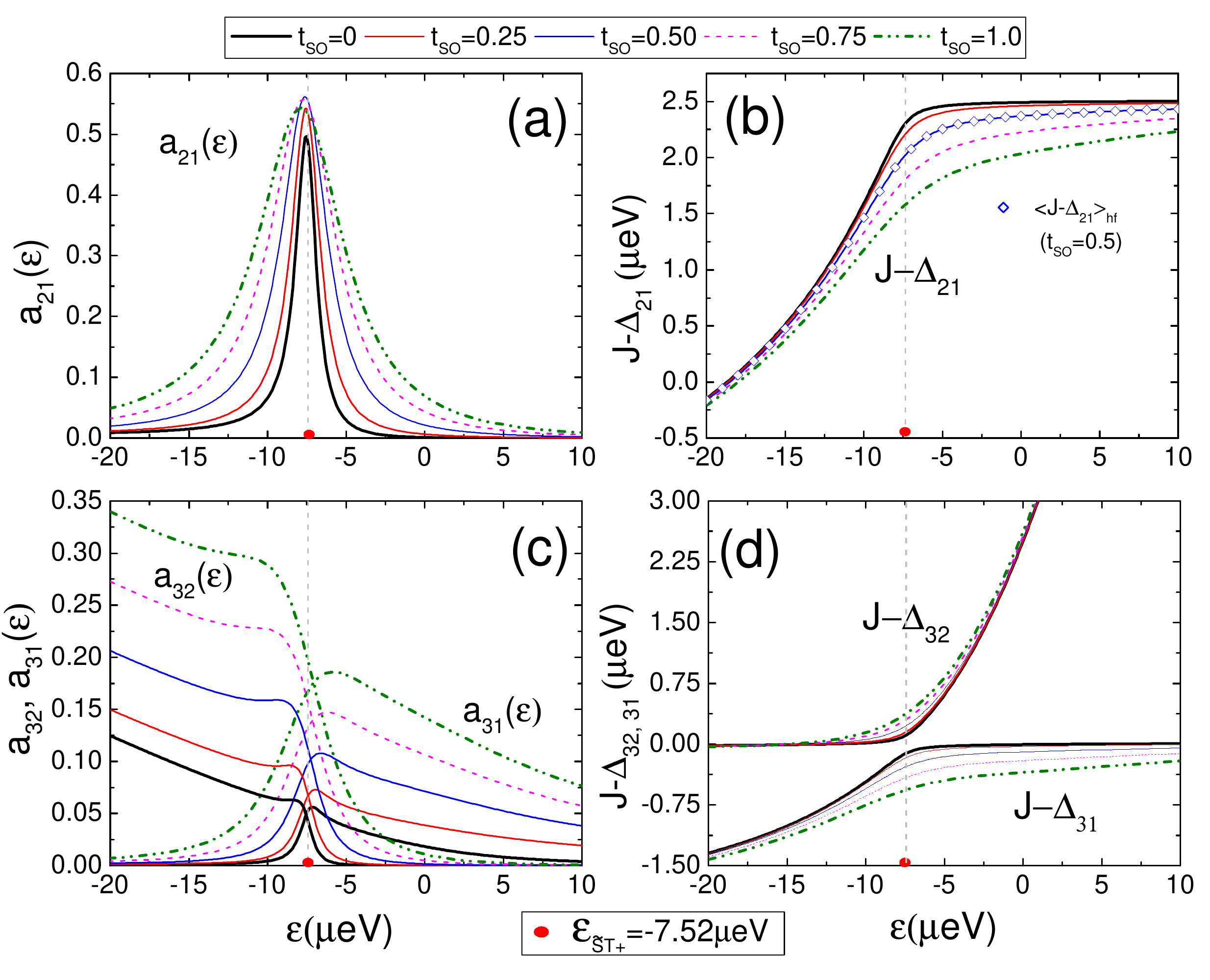}
    \caption{(Color online) Normalized Fourier amplitudes, $a_{ij}(\ep) = A_{ij}(\ep)/A_0$ (as given by Eqs.\ \ref{Amplitudes} and \ref{ZeroAmplitude}), and energy splittings, $J(\ep)-\Delta_{ij}(\ep)$ (relative to the exchange interaction), associated with the transitions $\ket{\lm_i} \rightarrow \ket{\lm_j}$ for different values of SOI non-spin conserving tunneling strength $t_{SO}$, indicated in top legend (in $\mu\text{eV}$). (a) Normalized amplitude $a_{21} = a_{\tl{S}T_+}$. (b) Energy splitting $\Dl_{21} = \Dl_{\tl{S}T_+}$ relative to $J$. Here, the scatter plot points (in blue) correspond to the mean value of energy splitting for $t_{SO}=0.5\mu \text{eV}$, considering a normal distribution of fluctuating hyperfine fields.  Notice no difference with fixed hyperfine field (solid blue curve) results.  (c) Normalized amplitudes $a_{32}$ and $a_{31}$. (d) Energy splittings $J-\Delta_{32}$ and $J- \Delta_{31}$. In all graphs, the $\tl{S}-T_+$ avoided crossing resonance is indicated by a red dot marked on the x-axis at $\ep_{\tl{S}T_+}=-7.52\mu\text{eV}$.}
    \label{AmplitudesFrequencies}
\end{figure*}

Let $A_{mn}$ denote the transition amplitude from the state $m$ into $n$, and $\Delta_{mn}$ the energy splitting between the instantaneous eigenstates $\lm_m$ and $\lm_n$, respectively. As shown in Fig.\ \ref{GenericSpectrumPulse}, we have labeled the detuning dependent eigenvalues in order of increasing energy, i.e. $\lm_3(\ep) > \lm_2(\ep) > \lm_1(\ep)$. It is clear that for $\ep \ll \ep_{\tl{S}T_+}$, the eigenstates approach the limits $\ket{\lm_1} \rightarrow \ket{T_+}$, $\ket{\lm_2} \rightarrow \ket{\tl{S}}$ and $\ket{\lm_3} \rightarrow \ket{T_0}$, while for $\ep \gg \ep_{\tl{S}T_+}$, the limits are $\ket{\lm_1} \rightarrow \ket{\tl{S}}$, $\ket{\lm_2} \rightarrow \ket{T_+}$ and $\ket{\lm_3} \rightarrow \ket{T_0}$. Figure \ref{AmplitudesFrequencies} shows the dependence on detuning and spin-orbit tunneling strength of the different $A_{mn}$ amplitudes normalized to the zero-frequency amplitude, $a_{mn}(\ep) = A_{mn}(\ep) / A_0(\ep)$; panels (b) and (d) also show the corresponding energy splitting with respect to the exchange energy, $J(\ep) - \Dl_{mn}(\ep)$. The Bloch-Feshbach projection allows us to obtain analytical expressions for the corresponding amplitudes and frequencies, as described in detail by Eqs.\ \ref{Amplitudes} and \ref{ZeroAmplitude} in Appendix \ref{FeshbachProjection}. In the following we describe their behavior as function of $t_{SO}$.

Figure \ref{AmplitudesFrequencies}(a) shows the transition amplitude $a_{21}(\ep)$ for different values of $t_{SO}$. Notice that $A_{21}=A_{\tl{S}T_+}$ for all $\ep$, so that it becomes maximal at the avoided crossing for $\ep_{\tl{S}T_+}$, where the maximum rate of spin-flip assisted tunneling occurs. Naturally, the width of the line-shape increases with increasing $t_{SO}$, enhancing the detuning range over which significant amount of mixing between the singlet $\tl{S}$ and the triplet $T_+$ state occurs. The line-shape has a slight asymmetry with respect to the position of the resonance. For $\ep < \ep_{\tl{S}T_+}$, both the amplitude and splitting change slowly and the mixing with the singlet $S(1,1)$ persists for a wide range of detunings. In contrast, for $\ep > \ep_{\tl{S}T_+}$, SOI assisted transitions occur via singlet-triplet coupling along the $S(0,2)$ component of the hybridized singlet $\tl{S}$, such that the amplitude decays faster as this component gets rapidly out of resonance with the triplet state $T_+$. Figure \ref{AmplitudesFrequencies}(b) shows the detuning dependence of the splitting, $\Dl_{21}(\ep) = \Dl_{\tl{S}T_+}$, characterizing the frequency of the $P_{\tl{S}}$ oscillations associated with the $\tl{S}-T_+$ transition for different $t_{SO}$ values ($\Dl_{21}$ increases with larger $t_{SO}$, as intuitively expected). General analytical expressions for the energy splittings in terms of $t_{SO}$ powers, are given in Appendix \ref{FeshbachProjection}, Eqs.\ \ref{STShift}-\ref{absW}. In addition, a full derivation of the frequency shifts associated to $\Delta_{\tl{S}T+}$ and $\Delta_{\tl{S}T_0}$ is given in Eqs.\ \ref{a1zero}-\ref{URExpansionSTPlus} and Eqs.\ \ref{a1zeroSTZero}-\ref{URExpansionSTZero}, respectively. The largest increase in each case ($\sim 0.7\mu\text{eV}$) for $t_{SO} \simeq 1\mu\text{eV}$ is observed near the vicinity of $\ep_{\tl{S}T_+} \simeq -7.0\mu\text{eV}$. Beyond the crossover region, SOI continues to play a significant role in the state dynamics as the energy shifts continue to be appreciable, a behavior that is consistent with that of $a_{21}(\ep)$.

Figure \ref{AmplitudesFrequencies}(c) shows the detuning dependence and SOI effects on the amplitudes $a_{32}(\ep)$ and $a_{31}(\ep)$, which contain information relevant to the exchange driven singlet-triplet transitions, as well as the much weaker triplet-triplet transitions. Notice that $\lm_1$ and $\lm_2$ switch character at $\ep_{\tl{S}T_+}$, as reflected in the $a_{32}$ and $a_{31}$ amplitudes. For $\ep < \ep_{\tl{S}T_+}$, $a_{32} \rightarrow a_{\tl{S}T_0}$ and the amplitude increases as the system enters the qubit subspace $S-T_0$, where exchange mediated processes dominate; the amplitude is larger as $t_{SO}$ increases. As the detuning approaches $\ep_{\tl{S}T_+}$, the amplitude decays with a slight revival just before reaching the point of closest approach at the avoided crossing. Beyond this point, for $\ep > \ep_{\tl{S}T_+}$, the system leaves the exchange-driven qubit subspace. Here, $a_{32} \rightarrow a_{T_0T_+}$ becomes the amplitude corresponding to triple-triplet transitions, and decays much faster beyond the crossover region. On the other hand,  it is clear that $a_{31} (\rightarrow a_{\tl{S}T_0})$ decays at a much slower rate with increasing detuning. This represents $a_{\tl{S}T_0}$ having a larger $S(0,2)$ component in $\tl{S}$. The singlet-triplet coupling enabling this transition is SOI, via non-spin conserving tunneling which couples all triplets to the outgoing singlet.

Correspondingly, Fig.\ \ref{AmplitudesFrequencies}(d) shows the energy splittings $\Dl_{31}(\ep)$ and $\Dl_{32}(\ep)$, again relative to the exchange interaction term, $J$. For $\ep < \ep_{\tl{S}T_+}$, it is clear that $\Delta_{32} \rightarrow \Delta_{\tl{S}T_0} \simeq J(\ep)$ and SOI-induced frequency shifts are hard to resolve in this limit. For $\Delta_{31} \rightarrow  \Delta_{T_0T_+}$, SOI effects are slightly more evident in the energy splittings as $t_{SO}$ increases, even at detuning values far from the $\tl{S}-T_+$ resonance. Yet, the corresponding transitions have a very low amplitude, $a_{31}$, as shown in Fig.\ \ref{AmplitudesFrequencies}(c). For $\ep > \ep_{\tl{S}T_+}$, however, the component $\Delta_{31}\rightarrow\Delta_{\tl{S}T_0}$ is the one that better resolves the frequency shifts accompanied with a significative increase in the corresponding amplitude $a_{31}$. It is evident in all the figures that SOI effects are amplified near the vicinity of $\ep_{\tl{S}T_+}$. In this crossover region, the $\tl{S}-T_0$ component of both splittings, $\Delta_{31}$ and $\Delta_{32}$, shows the largest effects with increasing $t_{SO}$ values, although both splittings and amplitudes are typically much smaller than those shown in the $\tl{S}-T_+$ component of $a_{21}$ and $\Delta_{21}$, see Figs.\ \ref{AmplitudesFrequencies}(a) and (b).

\section{Hyperfine field fluctuations and electrical noise effects}
\label{Noise}

In contrast to SOI, which is essentially static in a given structure, the hyperfine interaction has a dynamic and random character, as the nuclear magnetic field changes in time. Therefore, in-between consecutive detuning sweeps, required to enhance the signal to noise ratio, fluctuations in the hyperfine field (even as they may occur over a ms scale or longer) are expected to result in a slightly different frequency of Rabi oscillations involving $\tl{S}$, $T_0$, and $T_+$. Consequently, there is an associated experimental uncertainty in the estimation of the frequency shifts shown in Fig.\ \ref{AmplitudesFrequencies}(b). To estimate the error associated with variations in the nuclear magnetic field gradient, $dB$, we have calculated the mean value of the energy splitting, $\Delta_{21}$, as shown by the scatter plot symbols in Fig.\ \ref{AmplitudesFrequencies}(b). The mean splitting averaged over a distribution of nuclear hyperfine
fields $g_{\rm hyp}$, is given by
\begin{equation}
\langle J-\Delta_{21}\rangle_{\rm hf} = \int (J-\Delta_{21}) \, g_{\rm hf}(dB) \, d(dB) \, ,
\end{equation}
which averages over the variation of the nuclear magnetic field gradients, assumed to be given by a Gaussian distribution of width $\sigma_{\rm hf}$.   Here,
$g_{\rm hf}(x)=(1/\sigma_{\rm hf}\sqrt{2\pi})\exp(-x^2/2\sigma_{\rm hf}^2)$.
One would have
expected that the mean values near the vicinity of the $\tl{S}-T_+$ avoided crossing would exhibit the largest deviation. Yet we see that $\abs{\Delta_{21}-\langle\Delta_{21}\rangle_{\rm hf}} \simeq 0$ over the entire $\varepsilon$ range shown for $t_{SO}=0.5\mu\text{eV}$, when the value of $\sigma_{\rm hf} = 0.125\mu$eV, typical in experiments, is used. \cite{Nichol,Martins} These results suggest that the value of $t_{SO}$ obtained from the $P_{\tl{S}}$ frequency analysis would be only minimally affected by the hyperfine field fluctuation values.  Clearly, larger $\sigma_{\rm hf}$ values would produce a larger error in the measured $t_{SO}$.

On the other hand, recent experiments have shown that electrical noise effects should be taken into account when the dynamics involves a voltage difference (detuning) between left and right quantum dots, as in our case.\cite{Martins} Such noise in the detuning voltages results in noise in the coherent exchange $J$ which affects the robustness of the system oscillations.
 To estimate the role of electrical noise effects, we calculate the singlet return probability as
\begin{equation}
\langle P_{\tl{S}} \rangle_{el} = \int P_{\tl{S}}(\ep,\tau_R) \, g(\ep-\ep_R) \, d\ep \, ,
\label{NoisePS}
\end{equation}
where $g(\ep-\ep_R)$ is a Gaussian distribution of width $\sigma$ centered on $\ep_R$ given by
\begin{equation}
g(\ep-\ep_R) = \frac{1}{\sqrt{2\pi}\sigma} \exp\lpa{-\frac{(\ep-\ep_R)^2}{2\sigma^2}} \, .
\label{Gaussian}
\end{equation}
Note that integration interval in Eq.\ \ref{NoisePS} is over the entire range, but a number $(> 6)$ of
$\sigma$-values results in fully converged results.

\begin{figure}[h]
\includegraphics[width=1.0 \columnwidth]{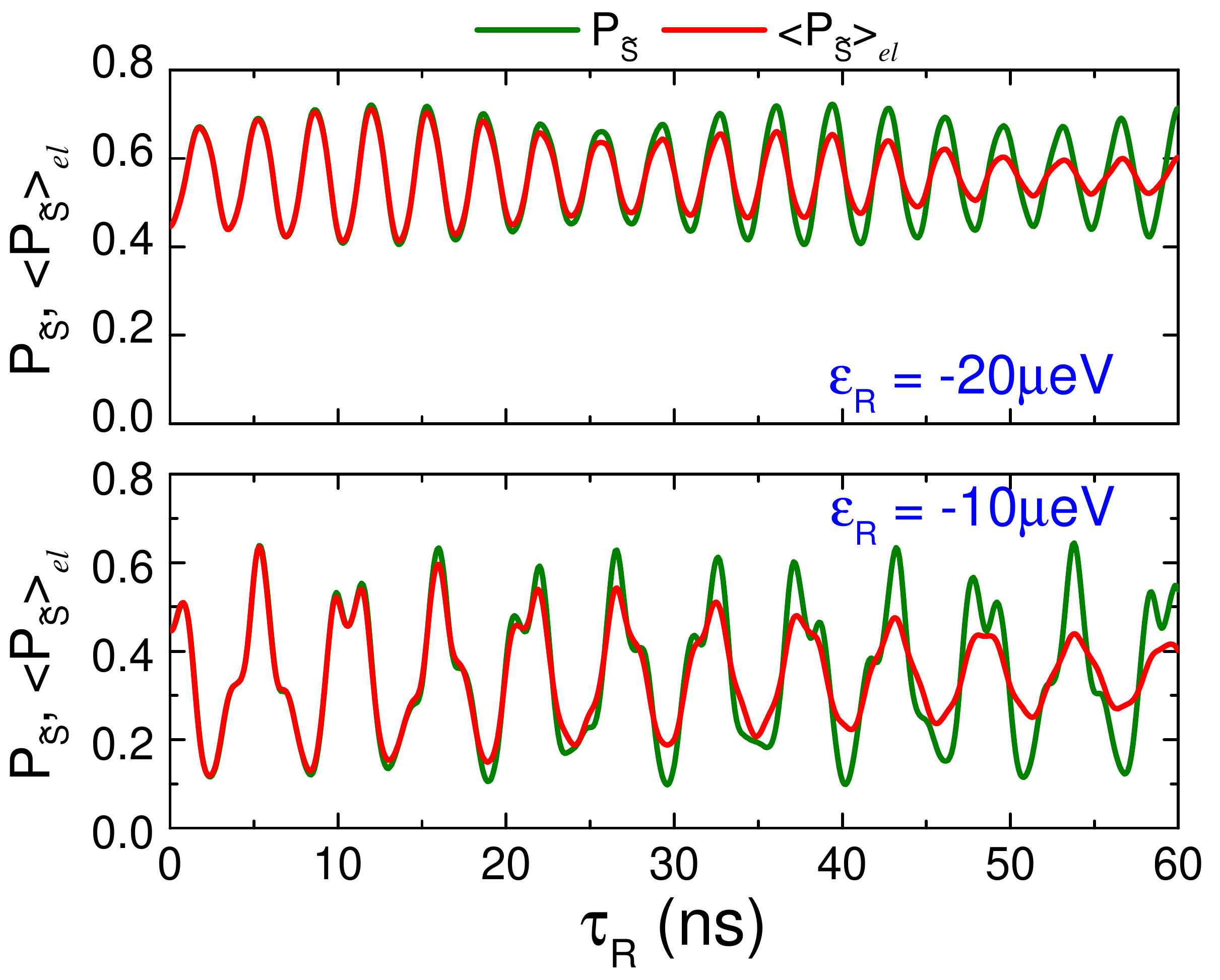}
\caption{(Color online) Singlet return probability $\langle P_{\tl{S}} \rangle_{el}$ ($P_{\tl{S}}$ ) in the presence (absence) of electrical noise effects as function of $\tau_R$ for $t_{SO}=0.8 \mu$eV. Noise effects in red solid curves are described by a Gaussian distribution with $\sigma=0.36\mu$eV as described in Eqs.\ \ref{NoisePS} and \ref{Gaussian}. Top panel: Return probability for $\ep_R = -20\mu$eV. Bottom panel: Return probability for $\ep_R = -10\mu$eV. The green solid curves in top (bottom) panels are same curves shown at the bottom panels of Figs.\ \ref{SpinOrbitSignatures}(b) and \ref{SpinOrbitSignatures}(d), respectively, and serve as comparison.}
\label{ChargeNoisePS}
\end{figure}

A realistic width of the distribution is obtained from recent experiments.\cite{Martins}
Martins {\em et al}. report an effective electrical gate noise of $\sigma_{el} = 0.18$mV and induced exchange oscillations  $\Delta J =116$MHz ($0.48\mu$eV) corresponding to a detuning voltage change of $2.5$mV. Using the relation $J=\frac{1}{2} \lpa{\ep + \sqrt{\ep^2 + 4t_0^2}}$, we obtain $\Delta J=0.96\mu$eV for a detuning change $\Delta \ep=10\mu$eV, which yields a scaling factor of 0.5 mV/$\mu$eV.\@ This translates $\sigma_{el}$ into our corresponding width of the electrical noise distribution as $\sigma=0.36\mu$eV. It is clear that $\sigma$ may be different in other experiments, but this value provides us with a realistic estimate to evaluate the effect of electrical noise on the SOI signatures we study.  To illustrate the role of electrical noise, we focus on the singlet return probability $P_{\tl{S}}$ shown in Fig.\ \ref{SpinOrbitSignatures}(b) and (d). The corresponding results for $\langle P_{\tl{S}} \rangle_{el}$ are shown in Fig.\ \ref{ChargeNoisePS},  for the two values of detuning $\ep_R=-20\mu$eV and $\ep_R=-10\mu$eV; it is evident that electrical noise dampens the amplitude of the oscillations of $P_{\tl{S}}$ as expected, but not the frequency.   As long as the amplitude remains sufficiently large (measurable) throughout the time interval shown, it should then be possible to carry out the frequency analysis we propose in order to extract quantitative values of the spin-orbit interaction.  It is also clear that larger noise fluctuations would strongly suppressed the coherence oscillations and make this (or any) analysis difficult.

%

\section{Summary and conclusions}
\label{Conclusion}

We have studied the signatures of spin-orbit interaction on the spectrum and dynamics of singlet-triplet qubits defined in two-electron GaAs double quantum dots. By reconstructing the level-anticrossing spectrum of the system as function of the interdot voltage detuning, we characterized the Rabi flopping dynamics originating from singlet-triplet transitions within the $\tl{S}-T_0$ and $\tl{S}-T_+$ qubit subspaces. This characterization allowed us to obtain the return probability of the singlet state as one applies voltage detuning sweeps traversing the $\tl{S}-T_+$ anticrossing resonance. The return probability exhibits an oscillatory behavior with frequencies and Fourier amplitudes that are strongly modulated by the spin-orbit non-spin conserving tunneling strength, and are more visible for residence detunings, $\ep_R$, at which the system is allowed to evolve near the $\tl{S}-T_+$ resonance. Furthermore, when taking into account the effects of electrical noise during the sequence of detuning sweeps, the oscillations of the singlet return probability persist, although with an overall dampening of their amplitude over the time intervals considered.  However, as long as the noise is not too strong, the analysis of the oscillations would still yield estimates of the spin-orbit coupling in the system.

By projecting the Hamiltonian of the system onto a subspace spanned by the states relevant to the crossover region of the $\tl{S}-T_0$ and $\tl{S}-T_+$ qubits, we obtained comprehensive analytical expressions that yield the dependence of the corresponding transition amplitudes and Rabi frequency shifts as function of all coupling parameters. The obtained signatures are the result of the interplay between exchange interaction and non-spin conserving processes originating from SOI and the hyperfine interaction between electron spins and those of the GaAs host nuclei. Our findings provide further insights into SOI signatures that could be probed by one-shot readout experiments measuring the singlet state return probability following a rapid detuning sweep traversing the $\tl{S}-T_+$ resonance.

An interesting avenue for further research, in the context of the present work, is the design of coherent control pulses aimed at reducing noise effects in the symmetric configuration (zero detuning), where variations in the exchange interaction are completely determined by charge tunneling.~\cite{Martins,Reed}

\acknowledgments
J.E.R. acknowledges the support of Consejo Nacional de Ciencia y Tecnolog\'ia (CONACYT) and useful discussions with J.E. Drut. E.C. acknowledges the support of DGAPA-UNAM PAPIIT project IN112012. S.E.U. acknowledges the support of the National Science Foundation under Grant DMR 1508325. We thank the Ohio Supercomputer Center for computational resources.


\appendix

\section{Analytical estimation of amplitudes and frequency shifts}
\label{FeshbachProjection}

\subsection{The Bloch-Feshbach projection method}
\label{BFMethod}

The level anticrossing signature between the hybridized singlet $\tl{S}$ and the triplet $T_+$ points to the onset of a non-trivial quantum coherent interaction mediated by the nuclear hyperfine interaction and spin-orbit coupling. In particular, the dependence of this interaction on the couplings strengths $\abs{\vec{t}}$ and $\abs{d\vec{B}}$ cannot be directly obtained from the off-diagonal matrix elements of the Hamiltonian in Eq.\ \ref{TildeHamiltonian}, nor from the level diagram shown in Fig.\ \ref{BareLevelDiagram}. However, the physics can be revealed by an effective Hamiltonian, $\mcl{H}_\text{eff}$, resulting from the projection of the full Hamiltonian onto a reduced sector of the Hilbert space containing eigenvectors relevant to the anticrossing region, with eigenvalues matching exactly those of the full Hamiltonian. To this end, we employ a non-perturbative procedure based on the Bloch-Feshbach projection operator formalism.~\cite{BlochFeshbach1, BlochFeshbach2, Cohen}

Let us consider a closed quantum system with the Hamiltonian given by Eq.\ \ref{TildeHamiltonian}. The Hamiltonian can be separated into two parts, $\tlm{H}=\tlm{H}_0 + V$, where $\tlm{H}_0$ is the diagonal part, and $V$ is the part that contains the interactions that dress the bare spectrum of $\tlm{H}_0$. Let $\mathcal{P}$ be the relevant subspace spanned by the states that give rise to an avoided crossing resonance. Similarly, let $P$ and $Q=1-P$ be projector operators onto and outside of $\mcl{P}$, respectively. The effective Hamiltonian is given by
\begin{equation}
\tlm{H}_{\text{eff}}(z)=P\tlm{H}_{0}P+PR(z)P \, ,
\label{HamZ}
\end{equation}
with $z=E \pm i\ep$, where $E$ and $\ep$ are the real and imaginary parts of the complex energy eigenvalue $z$. The first term of $\tilde{H}$ is the leading part of the Hamiltonian inside $\mathcal{P}$, with the second term containing the level shift operator,
\begin{equation}
R(z)=V+VQ[z-QH_{0}Q-QVQ]^{-1}V \, ,
\label{LevelShiftOperator}
\end{equation}
which is projected onto $\mathcal{P}$. The latter term can be seen as a Hamiltonian that permits the calculation of the energy level shifts with respect to the unperturbed levels. Allowing the Hamiltonian to depend on its eigenvalues $z$, makes the eigenvalue equation non-linear. Additionally, analytic continuation of the eigenvalues into the complex plane allows the definition of a non-Hermitian Hamiltonian that could incorporate dissipation processes taking place outside the relevant subspace, $\mathcal{P}$. Self-consistent solutions to the non-linear eigenvalue equation are used to obtain the eigenvalue spectrum in the vicinity of a level crossing and anticrossing. Near a level anticrossing (and in the absence of accidental degeneracies) there is a unique self-consistent solution of $z(\ep)$ for each value of the applied bias detuning $\ep$.

\subsection{Singlet-triplet transition amplitudes}
\label{TransitionAmplitudes}

To qualitatively evaluate the behavior of both the Fourier amplitudes and frequency shifts associated with $P_{\tl{S}}$ as function of $t_{SO}$, starting from Eq.\ \ref{TildeHamiltonian} we apply the Bloch-Feshbach projection method to obtain an effective Hamiltonian, $\mcl{H}_{\tl{S}T}$, which describes the system dynamics within the subspace spanned by the states $\lca{\ket{\tl{S}}, \ket{T_+}, \ket{T_0}}$, which are the relevant states to the dynamical processes taking place near the $\tl{S}-T_+$ resonance. We can construct the time evolution of the initial state of the system, $\ket{X(0)} = (\ket{\tl{S}} + \ket{T_0})/\sqrt{2}$, such that the time evolution of the singlet recovery is given by
\begin{eqnarray}
P_{\tl{S}}(t) &=& \abs{\ovr{\tl{S}}{X(t)}}^2 \\ \nonumber
          &=& A_0 + 2~\text{Re}~(A_{32}e^{\frac{-i}{\hbar} \Dl_{32} t}  + \sum_{m=2}^3 A_{1m}e^{\frac{-i}{\hbar} \Dl_{1m} t}) \, .
\label{FullAmplitude}
\end{eqnarray}
Here, $\Dl_{mn}$ are the transition frequencies between instantaneous eigenstates $\ket{\lm_m}$ and $\ket{\lm_n}$ of $\mcl{H}_{\tl{S}T}$, with corresponding amplitudes given by
\begin{equation}
A_{mn} = \frac{1}{2}(\abs{G_{1m}}^2+G_{1m}G_{3m}^*)(\abs{G_{1n}}^2 + G_{1n}^*G_{3n}) \, ,
\label{Amplitudes}
\end{equation}
where $G_{mn}$ are the matrix elements of the unitary operator, $\mcl{G}$, having columns formed by the eigenvectors, $\ket{\lm_m}$, represented in the basis $\lca{\ket{\tl{S}}, \ket{T_+}, \ket{T_0}}$.
Likewise, the zero-frequency amplitude is given by
\begin{equation}
A_0 = \sum_{m=1}^3 \frac{1}{2} \abs{G_{1m}}^2 \lpa{\abs{G_{1m}}^2+\abs{G_{3m}}^2 +2~\text{Re}~G_{1m}^*G_{3m}} \, .
\label{ZeroAmplitude}
\end{equation}

Equations\ \ref{Amplitudes} and \ref{ZeroAmplitude} allow to calculate the dependence on detuning and spin-orbit tunneling strength of the different $A_{mn}$ amplitudes normalized to the zero-frequency amplitude, $a_{mn}(\ep) = A_{mn}(\ep) / A_0(\ep)$, as discussed in the text and in Fig.\ \ref{AmplitudesFrequencies}.

\subsection{Frequency shifts associated with the singlet-triplet transitions}
\label{FrequencyShifts}

To obtain analytical estimates of the frequency shifts associated with transitions having predominantly $\tl{S}$-$T_+$ and $\tl{S}$-$T_0$ character, we adiabatically eliminate in each case the hybridized singlet $\tl{G}$ and triplet state $T_-$, while retaining their dynamical effects by including (to all orders) the resulting perturbative corrections to the matrix elements of the projected two-level Hamiltonian. In the relevant subspace $\mcl{P}$ spanned by either $\lca{\ket{\tl{S}}, \ket{T_+}}$ or $\lca{\ket{\tl{S}}, \ket{T_0}}$ the effective Hamiltonian is given by
\begin{equation}
\tlm{H}_{\text{eff}} = \left(
  \begin{array}{cc}
    E_1 & U_R + i U_I \\
    U_R + i U_I & E_2 \\
  \end{array}
\right) \, ,
\label{Hameff2x2}
\end{equation}
with eigenvalues given by
\eq{\lambda_1 = \frac{\Sigma - \sqrt{\delta^2 + 4\abs{W}^2}}{2}}
\begin{equation}
\lambda_2 = \frac{\Sigma + \sqrt{\delta^2 + 4\abs{W}^2}}{2} \,
\end{equation}
where $\Sigma = E_2 + E_1$, $\delta = E_2 - E_1$ and $W = U_R+i U_I$. Correspondingly, the frequency shift associated with a singlet-triplet transition is given by
\begin{equation}
\frac{\lm_2 - \lm_1}{\hbar} = \frac{1}{\hbar} \sqrt{\delta^2 + 4\abs{W}^2}
\label{STShift}
\end{equation}

In order to characterize the frequency dependence of the SOI tunneling strengths, we expand both $\dl$ and $\abs{W}$ in power series of $t_{SO}$, i.e.  $\delta = A_0(dB,..) + A_1(dB,..)t_{SO} + A_2(dB,..)t_{SO}^2 + ...$ The coefficients $A_n(dB,..)$ are functions of the remaining Hamiltonian parameters, in particular $dB$, which competes with $t_{SO}$. Therefore, we also expand $A_n$ in powers of $dB$ up to second order, $A_n = \al^{n}_0 + \al^{n}_1 dB + \al^{n}_2 dB^2 + ..$. For our two-level system projection, the coefficients multiplying odd powers of $dB$ and $t_{SO}$ vanish, i.e. $A_1 = A_3 = .. = 0$, $\al^{n}_1 = \al^{n}_3 = .. = 0$. Therefore
\begin{equation}
\delta = \al^{0}_0 + \al^{0}_2 dB^2 +(\al^{2}_0 + \al^{2}_2 dB^2)t_{SO}^2 \, ,
\label{DeltaExpansion}
\end{equation}

The complex off-diagonal coupling, $U= U_R + i U_I$, is expressed in polar form, i.e. $ U = \abs{W} e^{i\phi}$, where
\begin{equation}
\phi = \arctan{\frac{U_I}{U_R}} \, ,
\label{PhiAngle}
\end{equation}
such that
\begin{equation}
\abs{W} = U_R \sec{\phi} \, .
\label{absW}
\end{equation}

\subsection{$\tl{S}-T_+$ frequency shifts}
\label{STplusShifts}

After projecting out the states $\tl{G}$, $T_-$ and $T_0$, the coefficients in Eq.\ \ref{DeltaExpansion} are given by,
\eq{\al^{0}_0 = J + B_z  \, , \label{a1zero}}
\eq{\al^{0}_2 = -\frac{1}{J^2+t_0^2} \left( \frac{J^2}{J-\ep -z}+  \frac{t_0^2(B_z + 2z)}{z(B_z+z)} \right) \, , \label{a102}}
\eq{\al^{2}_0 = -\frac{1}{J^2+t_0^2} \left( \frac{t_0^2}{J-\ep -z} + \frac{J^2(B_z + 2z)}{z(B_z+z)}\right) \, .\label{a120}}
For the range of parameters considered here, $\abs{\al^{2}_2(\ep)} \ll \abs{\al^{2}_0(\ep)}$, so this latter coefficient can be neglected in the calculations.

Now, the phase associated with the off-diagonal coupling in Eq.\ \ref{PhiAngle} is given by
\begin{equation}
\phi = \arctan{\left(\frac{dB t_0 + J t_{SO}}{dB t_0 - J t_{SO}}\right)} \, ,
\label{PhiAngleSTPlus}
\end{equation}
while the power series expansion of $U_R$ in terms of $t_{SO}$ up to second order is given by,
\begin{equation}
U_R = -\frac{1}{\sqrt{2}} \frac{(dBt_0-J t_{SO})}{\sqrt{J^2 + t_0^2}} + \frac{1}{\sqrt{2}} \frac{ \frac{(B_z+2z)dB t_0^3}{(J-\ep-z)(B_z+z)z} }{\sqh{(J^2+t_0^2)}} t_{SO}^2 \, .
\label{URExpansionSTPlus}
\end{equation}
Substitution of these expressions in Eqs.\ \ref{DeltaExpansion}, \ref{PhiAngle}, and \ref{absW}, allows the explicit evaluation of the frequency shift, $\Delta_{21} = \Delta_{\tl{S}T_+}$ in Fig.\ \ref{AmplitudesFrequencies}(b), as function of both $t_{SO}$, $dB$, and the remaining coupling parameters.

\subsection{$\tl{S}-T_0$ frequency shifts}
\label{STZeroShifts}

For the most part, the shift of the splitting $\Delta_{32} \rightarrow \Delta_{\tl{S}T_0}$ with $t_{SO}$ corresponds essentially to SOI corrections to the exchange energy, $J$, which are in general much smaller in comparison to that of $\Delta_{\tl{S}T_+}$. Following the procedure outlined in the previous subsection, here we project out the states $\tl{G}$, $T_-$ and $T_+$. In this case, the coefficients in Eq.\ \ref{DeltaExpansion} are given by,
\eq{\al^{0}_0 = J  \, , \label{a1zeroSTZero}}
\eq{\al^{0}_2 = -\frac{1}{J^2+t_0^2} \left( \frac{J^2}{J-\ep -z} -  \frac{2t_0^2 z}{(B_z^2-z^2)} \right) \, , \label{a102STZero}}
\eq{\al^{2}_0 = -\frac{1}{J^2+t_0^2} \left( \frac{t_0^2}{J-\ep -z} -  \frac{2J^2 z}{(B_z^2-z^2)} \right) \, .\label{a120STZero}}
As before, $\abs{\al^{2}_2(\ep)} \ll \abs{\al^{2}_0(\ep)}$, so that coefficient can also be neglected. The phase associated with the off-diagonal coupling in Eq.\ \ref{PhiAngle} is given by
\begin{equation}
\phi = \arctan{\left(\frac{J t_{SO}}{dB t_0}\right)} \, ,
\label{PhiAngleSTZero}
\end{equation}
while the power series expansion of $U_R$ in terms of $t_{SO}$ up to second order yields
\begin{equation}
U_R = \frac{dBt_0}{\sqrt{J^2 + t_0^2}} + \frac{ \frac{(2z)dB t_0^3}{(J-\ep-z)(B_z^2-z^2)} }{\sqh{(J^2+t_0^2)}} t_{SO}^2 \, .
\label{URExpansionSTZero}
\end{equation}
Finally, by substituting these expressions in Eqs.\ \ref{DeltaExpansion}, \ref{PhiAngle}, and \ref{absW}, one can explicitly evaluate the exchange driven frequency shift, $\Delta_{32}$, shown in Fig.\ \ref{AmplitudesFrequencies}(d), as function of both $t_{SO}$, $dB$, and the remaining coupling parameters.



\end{document}